\documentclass[twocolumn, prl, superscriptaddress,notitlepage]{revtex4-2}
\usepackage{graphicx}
\usepackage{dcolumn}
\usepackage{bm}
\usepackage[usenames,dvipsnames]{color}
\usepackage[most]{tcolorbox}
\usepackage{multirow}
\usepackage{gensymb}
\usepackage[normalem]{ulem}
\usepackage{comment}
\usepackage{hyperref}
\hypersetup{colorlinks,allcolors=blue}
\usepackage{amssymb}
\usepackage{pifont}
\usepackage{physics}
\usepackage{natbib}

\usepackage{color,soul}

\begin{document}

\title{Non-Local and Non-Markovian Effects of a Microscopic Two-Level Defect in Superconducting Quantum Circuits}

\newcommand{\BAQIS}{\affiliation{1}{Beijing Key Laboratory of Fault-Tolerant Quantum Computing, \\
Beijing Academy of Quantum Information Sciences, Beijing 100193, China}}
\newcommand{\HFNL}{\affiliation{2}{Hefei National Laboratory, Hefei 230088, China}}
\newcommand{\IOP}{\affiliation{3}{Beijing National Laboratory for Condensed Matter Physics,\\
Institute of Physics, Chinese Academy of Science, Beijing 100190, China}}
\newcommand{\UCAS}{\affiliation{4}{University of Chinese Academy of Sciences, Beijing 101408, China}}
\newcommand{\THUCS}{\affiliation{5}{Department of Computer Science and Technology, Tsinghua University, Beijing 100084, China}}

\author{Yang Gao}
\thanks{These authors have contributed equally to this work.}
\affiliation{\BAQIS}
\affiliation{\IOP}
\affiliation{\UCAS}

\author{Yujia Zhang}
\thanks{These authors have contributed equally to this work.}
\affiliation{\BAQIS}
\affiliation{\IOP}
\affiliation{\UCAS}

\author{Huikai Xu}
\email{xuhk@baqis.ac.cn}
\affiliation{\BAQIS}

\author{Pan Shi}
\affiliation{\BAQIS}

\author{Feiyu Li}
\affiliation{\BAQIS}
\affiliation{\IOP}
\affiliation{\UCAS}

\author{Yaqing Feng}
\affiliation{\BAQIS}

\author{Weijie Sun}
\affiliation{\BAQIS}

\author{Jiayu Ding}
\affiliation{\BAQIS}

\author{Yang Liu}
\affiliation{\BAQIS}

\author{He Wang}
\affiliation{\BAQIS}

\author{Ruixia Wang}
\affiliation{\BAQIS}

\author{Zhen Yang}
\affiliation{\BAQIS}

\author{Yirong Jin}
\affiliation{\BAQIS}

\author{Haifeng Yu}
\email{hfyu@baqis.ac.cn}
\affiliation{\BAQIS}
\affiliation{\HFNL}

\author{Fei Yan}
\email{yanfei@baqis.ac.cn}
\affiliation{\BAQIS}

\begin{abstract}
Microscopic two-level systems (TLS) --- ubiquitous atomic-scale defects in solid-state quantum devices --- are a dominant source of qubit decoherence, yet their role is often considered local and short-memoried. Here, we report the observation of a coherent TLS that couples simultaneously to two spatially distant superconducting qubits. The TLS is identified to reside within the tunable coupler linking the qubits, enabling controllability of the TLS-qubit coupling strength via coupler frequency --- a capability absent in earlier studies. This tunability allows us to systematically probe how TLS distorts qubit dynamics, revisiting the decoherence model in the presence of non-Markovian TLS dephasing noise. This is corroborated by the reconstructed $1/f$ noise spectrum of TLS frequency fluctuation spanning more than ten orders of magnitude (0.1\,mHz -- 1\,MHz) that reveals discrete fluctuator signatures. Quantum process tomography further unveils TLS-induced correlated qubit dynamics, highlighting the long-lived TLS as an effective source of non-Markovianity. Our findings expose a previously overlooked interaction mechanism in scalable quantum architectures: defects embedded in coupling elements can simultaneously affect multiple qubits with variable impact. Beyond immediate implications for system characterization and calibration, this situation provides a powerful testbed for studying defect-driven quantum dynamics, refining error suppression strategies, and advancing architecture design for scalable quantum technologies.
\end{abstract}

\maketitle

\section{Introduction}

The pursuit of fault-tolerant quantum computing requires understanding not only qubit properties but also their interactions with the environment. For superconducting circuits---a leading platform for quantum error correction\cite{google_quantum_ai_and_collaborators_quantum_2025,krinner_realizing_2022,zhao_realization_2022,wang2025demonstrationlowoverheadquantumerror}---two-level systems (TLS) represent the most pervasive sources of decoherence and control errors~\cite{muller_towards_2019,siddiqi_review_coherence_2021,megrant_chenyu_scaling_2025,simmonds_decoherence_2004,martinis_decoherence_2005,ithier_decoherence_2005,wilen_correlated_2021,schlor_correlating_2019,cho_simulating_2023,lisenfeld_enhancing_2023}.  These atomic-scale defects, distributed randomly in both space and frequency with temporal fluctuations\cite{lisenfeld_observation_2015,klimov_fluctuations_2018,bilmes2020resolving,burnett_decoherence_2019,carroll_dynamics_2022,chen_phonon_2024,colao_zanuz_mitigating_2025,chen_scalable_TLS_control_2025}, introduce uncontrolled degrees of freedom that hybridize with qubits near resonance, transforming simple qubit operations into complex dynamics that severely degrade quantum operations.
Besides efforts to reduce TLS densities and coupling strengths by improved fabrication~\cite{Crowley2023DisentanglingTa,Wolff2026StructuralControlTLS} and qubit design~\cite{Martinis2022TaperedWiring}, a prevailing mitigation strategy employs careful frequency planning---tuning qubit operating points to avoid resonant TLS interactions\cite{klimov_snake_2020,klimov_optimizing_2024}. This approach relies on a comprehensive understanding of how TLS interact with qubits. Most studies are based upon assumptions that TLS primarily interact locally with individual qubits at a fixed coupling strength, and that their decoherence mechanisms follow simple Markovian (memoryless) dynamics~\cite{barends_coherent_2013,spiecker_solomon_2024,shalibo_lifetime_2010,delben_control_2023}. However, growing evidence reveals that there are a diverse species of TLS including those leading to non-Markovian qubit dynamics~\cite{niu2019learningnonmarkovianquantumnoise,cleland_studying_2024,oda_sparse_2024,zhuang_non-markovian_2025,spiecker_two_level_2023,gravier_simulated_2025,nila2025_continuous_quantum_correction_markovian,odeh_non-markovian_2025_phonon,Grabovskij_2011}, where even a single defect can catastrophically disrupt qubit operation---a critical concern for scaling quantum processors.

Here, we report the identification of a coherent TLS simultaneously coupled to two spatially separated qubits via their tunable coupler that hosts the TLS. This reveals an unexplored interaction pathway where defects in coupling elements mediate non-local effects. Our platform enables continuous tuning of the TLS-qubit coupling strength through coupler frequency adjustment, permitting a systematic study of qubit decohering processes. Using noise spectroscopy techniques, we reconstruct the TLS frequency noise spectrum (0.1\,mHz--1\,MHz), observing discrete fluctuators contributing to a $1/f$-like profile. Quantum process tomography further reveals TLS-mediated correlated qubit dynamics, while numerical simulations quantify how TLS parameters scale two-qubit gate errors. These findings compel reevaluation of defect interactions in quantum circuits, with direct implications for error modeling, processor calibration, and scalable quantum architecture design.

\subsection{An atomic-scale defect simultaneously coupled to two distant qubits}

\begin{figure*}[tbp]
\centering
\includegraphics[width=0.9\textwidth]{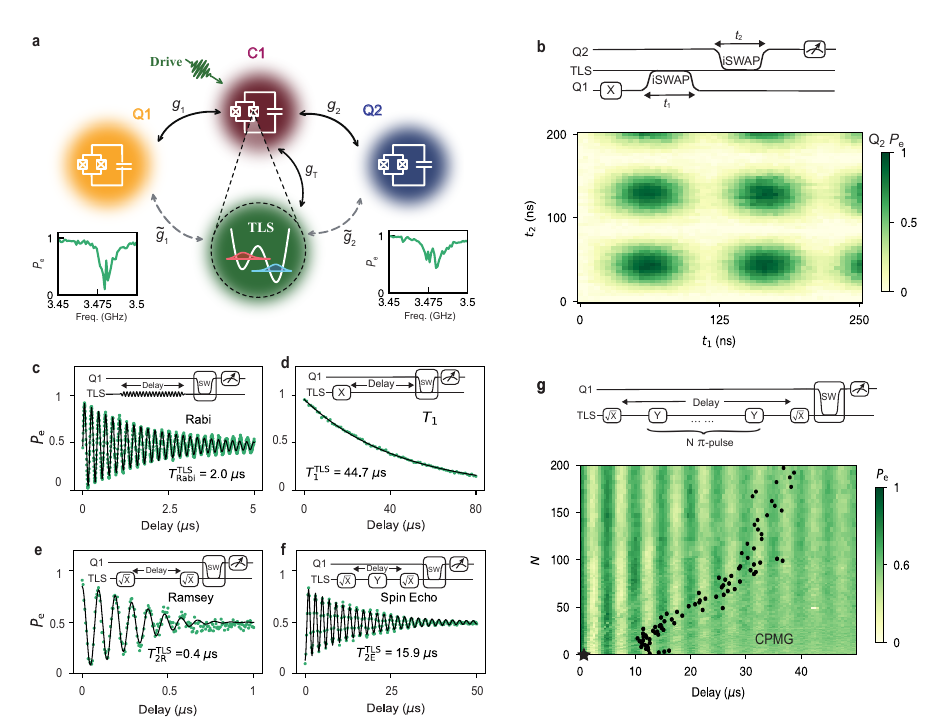}
\caption{
\textbf{Characterization of a microscopic defect residing in a tunable coupler.}
\textbf{a,} Experimental setup: two transmon qubits (separated by more than 1~mm in physical distance) coupled via a frequency-tunable transmon coupler hosting a TLS within a Josephson junction. Effective TLS-qubit couplings ($\tilde{g}_1$, $\tilde{g}_2$) arise from interactions mediated by the coupler ($g_\mathrm{T}$: TLS-coupler; $g_1$,$g_2$: qubit-coupler). Microwave pulses applied to the coupler directly drive the TLS via the cross-resonance effect.
\textbf{b,} Verification of the shared TLS: sequential iSWAP operations transferring an excitation from Q1 to TLS to Q2 by tuning the qubits into resonance with the TLS. Maximum Q2 excitation is achieved when pulse lengths $t_1$ and $t_2$ correspond to complete swaps. 
\textbf{c-f,} TLS coherence characterization via standard time-domain measurements: Rabi oscillations, energy relaxation, Ramsey dephasing, and spin-echo dephasing. Solid lines represent fits to exponential decay models (Gaussian decay for Ramsey). Insets are corresponding pulse sequences. TLS populations are measured after excitation swapping to Q1.
\textbf{g,} Enhanced TLS coherence via dynamical decoupling: measured CPMG decay curves (colormap) and extracted dephasing time constants (dots) plotted versus refocusing pulse count $N$, showing coherence revival. The star indicates the Ramsey ($N=0$) case for reference. In panels \textbf{e-g}, we update the phase of the final $\pi/2$ pulse linearly with the delay time to produce interference fringes for reliable coherence estimation.
}
\label{fig:characterization}
\end{figure*}

The subsystem studied here is part of a 72-qubit superconducting quantum processor with the tunable-coupling architecture~\cite{chen_efficient_2025}.
As illustrated in Fig.~\ref{fig:characterization}a, it consists of two frequency-tunable transmon qubits Q1 and Q2 (frequencies $\omega_{1}$ and $\omega_{2}$) coupled via a tunable transmon coupler C1 (frequency $\omega_{\rm C}$). We identify a coherent microscopic TLS, which is simultaneously observed in spectroscopic measurements of both qubits at frequency $\omega_{\rm T}/2\pi \approx 3.48~\text{GHz}$ (Fig.~\ref{fig:characterization}a). Usually, a superconducting qubit---macroscopic in size---would only be affected by defects that are physically local to its circuitry. However, this TLS, as we shall show later, is indeed strongly coupled to both qubits and turns out to be located within the coupler, which mediates the interaction between the TLS and qubits.
Starting from this hypothesis, we first perform a series of time-domain experiments to characterize the TLS.

To confirm that the TLS coherently interacts with both qubits, the coupler is first adjusted to the nominal bias where the two qubits are decoupled. The two qubits are then sequentially tuned into resonance with the TLS for controlled interactive durations. As a result, excitation can be controllably transferred from Q1 to Q2 (or vice versa) via TLS, resulting in the characteristic two-dimensional periodic pattern shown in Fig.~\ref{fig:characterization}b. 
This excludes the possibility of frequency coincidence of two independent TLS, one residing in each of the two qubits, confirming the presence of a coherent TLS shared by both qubits.
Additional cross-checks confirm that no similar spectral features appear in other qubits and the TLS is unaffected by any change in processor control bias with the exception of the coupler C1 frequency, which we shall show later. This excludes the possibility that the effect is due to another qubit or a common resonant mode. 
Furthermore, we observe random jumps in TLS frequency --- although very infrequently, which provides compelling evidence that it is an intrinsic defect rather than an artificial mode (see Supplementay Information Section VI).

Building on the hypothesis of strong TLS-coupler coupling, we utilize the cross-resonance effect \cite{lisenfeld_measuring_2010} to directly drive the TLS using microwave pulses applied to the coupler. This allows us to independently manipulate the TLS state and perform standard coherence characterization protocols - including Rabi oscillations, energy relaxation, Ramsey interferometry, spin echo, and CPMG dynamical decoupling sequences (Fig.~\ref{fig:characterization}c-g). Final TLS state readout is achieved through excitation transfer to Q1 via a calibrated $i$SWAP gate prior to measurement (see Supplementary Information Section VI and VII).

The TLS exhibits a notably long relaxation time of $T_1 = 44.7\pm1.3~\mu\text{s}$ and clear signatures of non-Markovian dephasing. Specifically, TLS coherence significantly improves under dynamical decoupling sequences, with coherence times extending from $T_{\rm 2R}\approx0.4~\mu\text{s}$ (Ramsey) to $T_{\rm 2E}\approx16~\mu\text{s}$ (echo) and further with increased refocusing pulse counts in CPMG sequences. This behavior---reminiscent of previous observations in superconducting qubits \cite{lisenfeld_decoherence_2016_CR_drive,bylander_dynamical_2011}---unambiguously demonstrates dominant low-frequency dephasing noise.


\subsection{Tunable TLS-qubit coupling and non-Markovian dynamics}

\begin{figure*}[htbp]
\centering
\includegraphics[width=1\textwidth]{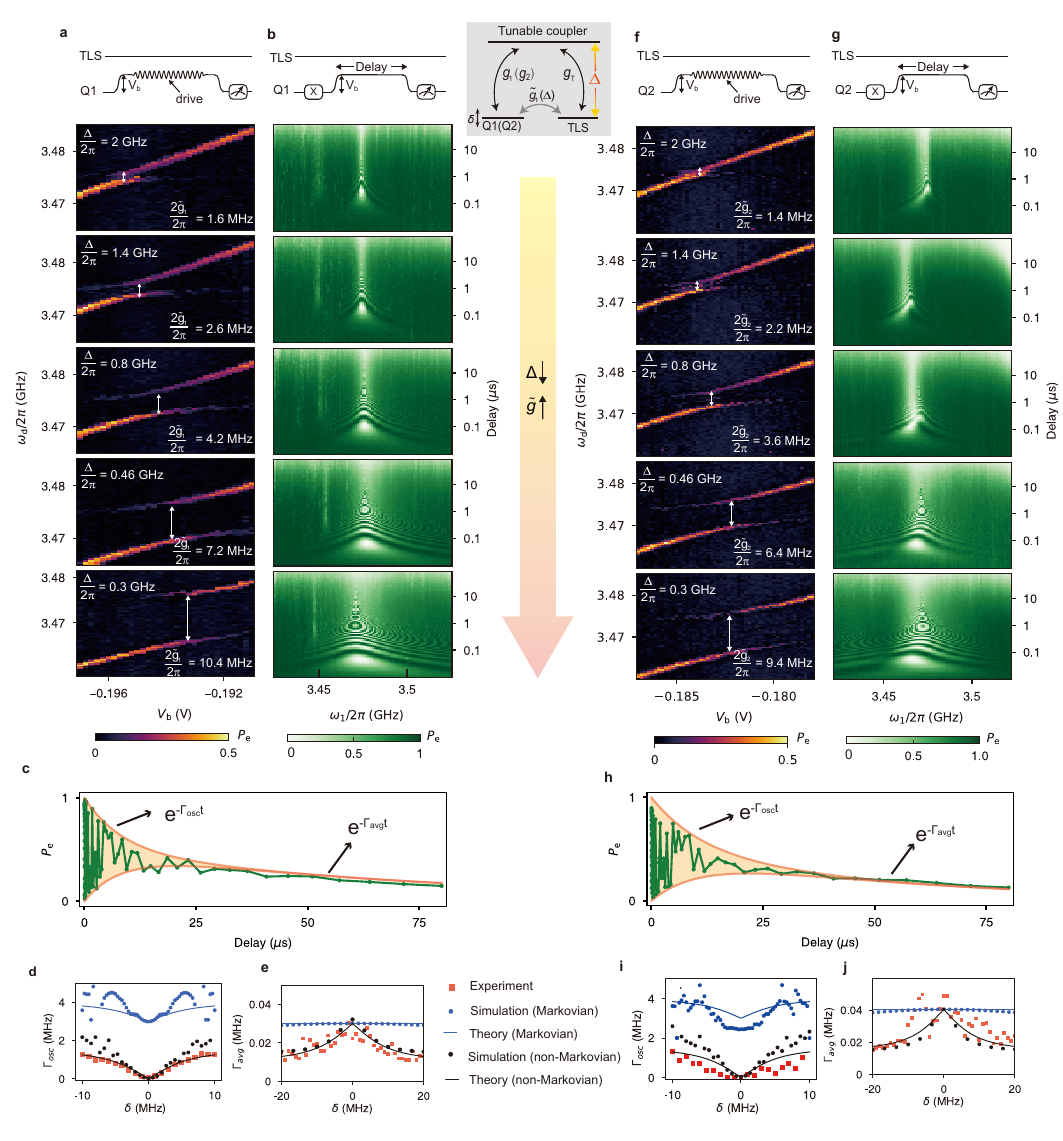}
\caption{
\textbf{Coupler-assisted TLS-qubit coupling.}
\textbf{a,} Spectroscopy of Q1 versus flux bias (pulse sequence illustrated above) for varying coupler-qubit detunings $\Delta$ while Q2 is biased far away from the relevant region. Avoided crossings in the spectrum (top to bottom) demonstrate enhanced $\tilde{g_1}$ as $\Delta$ decreases, accompanied by TLS frequency shift due to level repulsion from the coupler. The corresponding level diagram is illustrated in the inset in the top center.
\textbf{b,} Qubit dynamics under the $T_1$ relaxation protocol for varying qubit frequencies near the TLS resonance (pulse sequence illustrated above). Increased coupling accelerates qubit-TLS energy exchange and broadens the spectral interaction range (flux pulse rise/fall time: 2~ns). Note that the time axis is on a logarithmic scale.
\textbf{c,} An example $T_1$ trace measured at resonance. The dynamics include fast-decaying vacuum Rabi oscillations (decay rate $\Gamma_\mathrm{osc}$) upon a slower relaxation process (decay rate $\Gamma_\mathrm{avg}$). Note that the delay times are logarithmically sampled for experimental efficiency while still capturing the decaying envelop. A higher-resolution, linearly sampled case is shown in the Supplementary Information Section IX.
\textbf{d-e,} Extracted $\Gamma_\mathrm{osc}$ and $\Gamma_\mathrm{avg}$ versus qubit-TLS detuning compared with theory and numerical simulation by considering the dephasing noise being Markovian and non-Markovian. 
\textbf{f-j,} Corresponding results performed with Q2, confirming similarly tunable TLS coupling. The observed coupling strengths in panel \textbf{f} are consistently weaker for identical $\Delta$ values due to inherent asymmetries in the design of qubit-coupler couplings. Minor asymmetries of the chevron pattern in panel \textbf{g} are due to pulse distortions at short delays.
Misalignment of the resonance frequency of panel \textbf{g} is attributed to a random frequency jump of the TLS.
}
\label{fig:tunable}
\end{figure*}

We validate our system model (Fig.~\ref{fig:characterization}a) based on direct qubit-coupler ($g_{1}$,$g_{2}$) and TLS-coupler ($g_\mathrm{T}$) couplings by examining the effective qubit-TLS couplings $\tilde{g}_1$ and $\tilde{g}_2$ that emerge through virtual excitations of the coupler\cite{yan_tunable_2018}. In the dispersive coupling regime ($\Delta \gg g_{1}, g_{2}, g_\mathrm{T}$), the coupling strengths follow:
\begin{equation}
\tilde{g}_k \approx \frac{g_k g_{\rm T}}{\Delta},
\label{eq:g_tls_qubit}
\end{equation}
where $\Delta = \omega_{\rm C} - \omega_{\rm T} \approx \omega_{\rm C} - \omega_k$ ($k=1,2$) represents the coupler detuning from either the TLS or qubit (under the near-resonance condition $\omega_{\rm T} \approx \omega_k$). From spectroscopy, we extract the direct couplings: $g_1/2\pi = 70~\text{MHz}$ (Q1-coupler), $g_2/2\pi = 61~\text{MHz}$ (Q2-coupler), and $g_{\rm T} = 30~\text{MHz}$ (TLS-coupler; see Supplementary Information Section VI). Such a strong coupling suggests that the TLS is located in a tunnel barrier of the coupler's Josephson junctions \cite{martinis_decoherence_2005,simmonds_decoherence_2004,lisenfeld_observation_2015}, where the electric field density is high.
Reported TLS electric dipole moments $p$ are typically on the order of \(0.1\!-\!1~e\text{\AA}\), while local electric fields in junction tunnel barriers are generally estimated to be \(10^3~\mathrm{V/m}\), leading to an electric-dipole coupling strength up to a few tens of MHz \cite{muller_towards_2019}, consistent with the experimentally observed values.

Equation~\ref{eq:g_tls_qubit} reveals that $\tilde{g}_k$ becomes substantial (several MHz) when the coupler frequency approaches the qubit or TLS frequency (small $\Delta$), even while remaining in the dispersive regime ($\Delta \gg g_1, g_2, g_{\rm T}$). This suggests a previously overlooked mechanism: individual defects can significantly impact the performance of spatially separated qubits and cause non-local effects.

We experimentally confirm this tunable coupling capability by measuring the spectrum of Q1 as it sweeps through the TLS resonance (Fig.~\ref{fig:tunable}a). The observed avoided crossing shows a minimum gap, corresponding to twice the exchange coupling strength $2\tilde{g}_1$. As predicted, reducing the coupler frequency---and consequently $\Delta$---systematically increases the gap size. The measured coupling strengths agree well with theoretical values (see Supplementary Information VIII), validating our model of coupler-mediated interactions.

We characterize the qubit relaxation dynamics by tuning Q1 near the TLS resonance (Fig.~\ref{fig:tunable}b), where coherent excitation swapping produces distinctive chevron patterns. The vacuum Rabi oscillation rate reaches its minimum value $2\tilde{g}_1$ at zero qubit-TLS detuning ($\delta = \omega_1-\omega_\mathrm{T} = 0$) and can be controlled through the coupler detuning $\Delta$, with the interaction bandwidth scaling with $\tilde{g}_1$. In the strong coupling regime ($\tilde{g}_1 \gg \Gamma_1^\mathrm{TLS} = 1/T_1^\mathrm{TLS}$). The system exhibits dynamics characterized by two distinct timescales: (i) initial vacuum Rabi oscillations with a damping rate $\Gamma_\textrm{osc}$ upon (ii) a slower relaxation process with a rate $\Gamma_\textrm{avg}$ that typically persists longer. $\Gamma_\textrm{avg}$ can be conveniently extracted by fitting the slower decay after the oscillations vanishes. We obtain these rates across various detunings $\delta$ and show them in Fig.~\ref{fig:tunable}d-e.

The TLS dephasing affects the qubit relaxation dynamics---for example, it can dephase the initial vacuum Rabi oscillations. However, the extent critically depends on the noise's Markovian character~\cite{barends_coherent_2013,spiecker_solomon_2024}.
Notably, the qubit-TLS exchange interaction exhibits intrinsic robustness against quasistatic frequency fluctuations---a protection mechanism arising from field orthogonality. This robustness manifests itself in measured oscillation damping rates $\Gamma_\mathrm{osc}$ that consistently fall below Markovian predictions across all detunings $\delta$ (Fig.~\ref{fig:tunable}d).
The long-time relaxation rate follows $\Gamma_\mathrm{avg} = \Gamma_1^Q\cos^2\theta + \Gamma_1^\mathrm{TLS}\sin^2\theta$, where $\theta = \arctan(2\tilde{g}_1/\delta)$, $\Gamma_1^q$ is qubit relaxation rate. Its marked $\delta$-dependence also contradicts the nearly constant $\Gamma_\mathrm{avg}$ predicted by Markovian theory (see Supplementary Information Section VIII). 
Identical experiments with Q2 reproduce these results (Fig.~\ref{fig:tunable}f-j).

These results have significant implications for the design and operation of quantum processors:
\begin{itemize}
    \item \textbf{Variable coupling effects:} Standard characterization protocols performed at large coupler detunings may fail to capture the full impact of coupler-hosted TLS defects \cite{klimov_fluctuations_2018}. During two-qubit gates, when couplers are tuned near qubit frequencies to mediate interactions, TLS-qubit couplings are similarly enhanced, exacerbating gate performance degradation. Accurate modeling of these tunable couplings is crucial for both frequency planning and pulse optimization.

    \item \textbf{TLS characterization challenges:} The decoherence properties of TLS defects do not simply translate to qubit relaxation rates, as the coupling depends critically on details of the noise spectrum. Conventional $T_1$-based spectroscopy may overlook strongly dephasing defects, highlighting the need for comprehensive characterization methods such as state purity measurements~\cite{Auda_Zhu_purity_2024} and dynamical decoupling techniques.
    \item \textbf{Non-Markovian gate errors:} The observed initial oscillation damping from non-Markovian TLS dephasing introduces a distinct error mechanism for gate operations. For strong dephasing, it can induce significant coherent errors. 
\end{itemize}



\subsection{TLS Noise Spectroscopy} 

\begin{figure*}[!htbp]
    \centering
    \includegraphics[width = 1\textwidth]{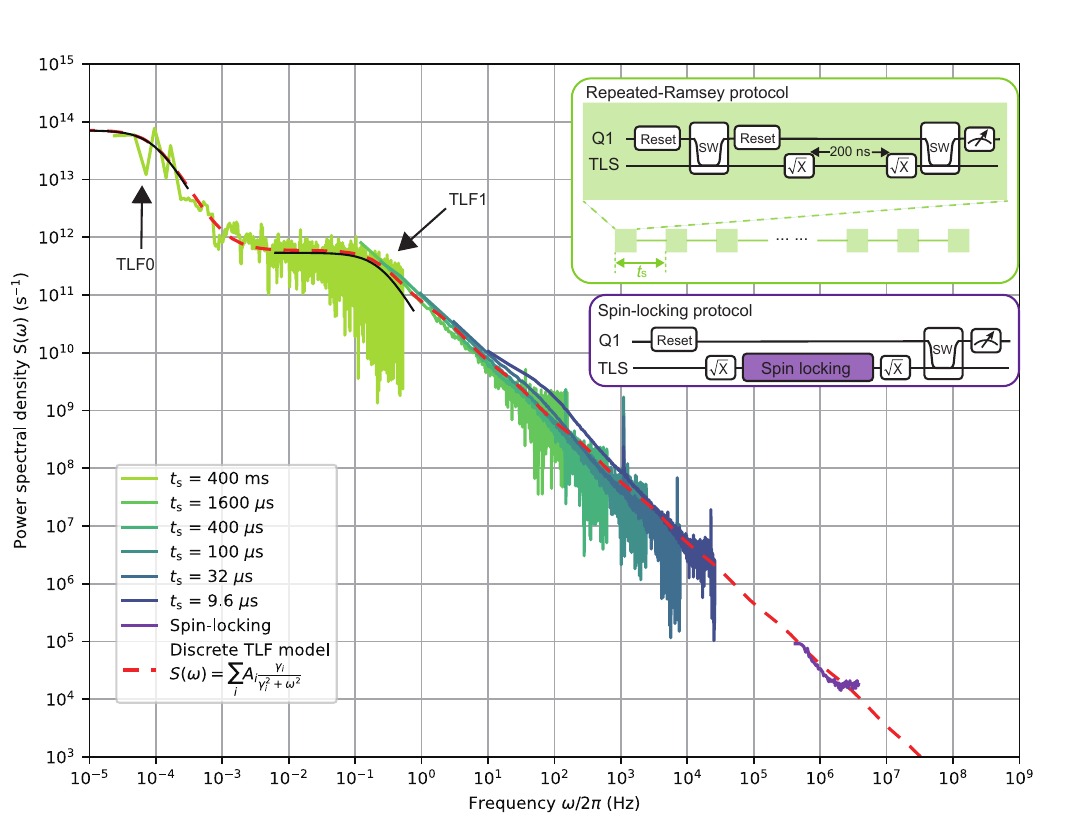}
    \caption{
    \textbf{Power spectral density of the TLS dephasing noise.}
    Multi-colored lines: noise PSD (0.1~mHz--10~kHz) derived from repeated Ramsey experiments of various sampling rates ($1/t_{\rm s}$).
    Purple dots: noise PSD ($\sim$1~MHz) derived from spin-locking experiments.
    Red dashed line: added-up noise spectrum of  several independent TLFs, with each Lorentzian spectrum plotted in black lines.
    Shown in the insets are the repeated-Ramsey and spin-locking protocols.
    We perform reset operations on both qubit and TLS when implementing the repeated-Ramsey protocol for high repetition rates ($>$100~kHz).
    The delay time between two $\sqrt{X}$ pulses is 0.2~$\mu$s.
    }
    \label{fig:tls_psd}
\end{figure*}

The coherence characterization results presented above reflect the convolution of the underlying noise spectrum with sequence-specific filter functions~\cite{bylander_dynamical_2011}. To reveal detailed information of the environmental noise affecting the TLS, we adapt qubit noise spectroscopy techniques~\cite{YAN_prb_2012_rep_ramsey,yan_rotating-frame_2013} to reconstruct the TLS frequency noise spectrum $S(\omega)$ across an extended range spanning 0.1~mHz to 1~MHz, covering ten decades of frequency.

For lower frequencies (0.1~mHz to 100~kHz), we implement a repeated-Ramsey protocol\cite{YAN_prb_2012_rep_ramsey,sank_flux_2012}. A Ramsey sequence with a fixed delay (200~ns) between $\sqrt{X}$ gates is repeated at a certain sampling interval $t_{\rm s}$ (Fig.~\ref{fig:tls_psd}, inset). For faster sampling rate, we implement successive reset operations on both qubit and TLS utilizing an on-chip dissipative mode \cite{ding_multipurpose_2025}(detailed in Supplementary Information Section VI). This way, the fluctuation of TLS frequency is encoded in the phase of Ramsey interference and hence the state projection probability. 
We employ cross-spectral analysis with interleaved sampling to suppress statistical noise (see Supplementary Information Section XI) and verify consistency by comparing results across different sampling intervals $t_{\rm s}$.

At higher frequencies (100~kHz to 1~MHz), we utilize the spin-locking or $T_{1\rho}$ relaxometry method~\cite{yan_rotating-frame_2013}, where the TLS relaxation during the spin-locking pulse --- the constant-amplitude $Y$ drive between $\sqrt{X}$ gates (Fig.~\ref{fig:tls_psd} --- is sensitive to $S(\omega)$ at the applied Rabi frequency $\Omega$. By systematically varying the drive amplitude, we map the noise spectrum across this frequency range, limited at low frequencies by decoherence from the dominant $1/f$ noise, and at high frequencies by available drive power.

The reconstructed spectrum reveals several important features. The overall profile follows a $1/f^{1.05}$ dependence, characteristic of distributed relaxation processes in solid-state systems\cite{paladino_1f_rmp_2014}. Below 1~Hz, we observe distinct bumps in the spectrum that may correspond to discrete two-level fluctuators (TLFs), each contributing a Lorentzian component, denoted as TLF0 and TLF1 in Fig.~\ref{fig:tls_psd}. This finding supports the Dutta-Horn model~\cite{dutta_low-frequency_1981}, where an ensemble of TLFs with properly distributed relaxation rates collectively produce $1/f$-type noise. 

Quantitative modeling using a set of 10 TLFs with switching rate ranging from 0.6~mHz to 0.2~GHz reproduces the measured spectrum. It should be emphasized that our intention is only to demonstrate that a relatively small number of fluctuators can already produce an approximately \(1/f\)-like spectrum. This does not imply that the actual number of TLFs in the system is necessarily so small; in practice, it may be substantially larger. Similarly, the switching-rate range used in the phenomenological model should not be interpreted as the actual physical distribution of TLF switching rates in the device. Integrating the modeled spectrum yields a predicted Ramsey dephasing time of $0.36\,\mu$s, in excellent agreement with experimental observations. 

Similar bump features and discrete fluctuator-like signatures have also been reported in superconducting quantum devices and related defect studies \cite{schlor_correlating_2019,burnett_decoherence_2019,Meibner18}. Evidence for low-frequency two-level fluctuators (TLFs) has also been observed in semiconductor quantum dots \cite{burkard2023semiconductor}. While $1/f$ charge noise has long been established as the dominant dephasing mechanism in semiconductor spin qubits, recent experiments have resolved the dynamics and temperature dependence of individual charge TLFs, with switching rates spanning 100~mHz to 100~Hz \cite{ye2024characterization,ye2025measuring}.

\subsection{Multipartite dynamics with TLS}

\begin{figure*}[!htbp]
    \centering
    \includegraphics[width = 1\textwidth]{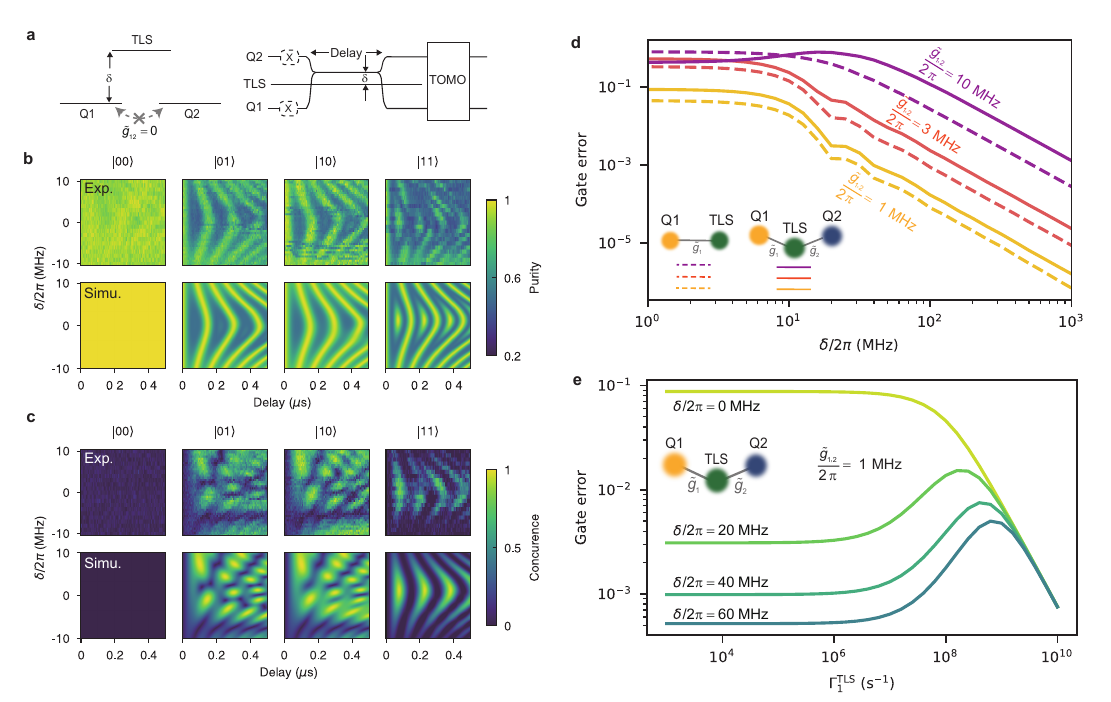}
    \caption{
    \textbf{Correlated qubit dynamics induced by TLS.} 
    \textbf{a,} Pulse sequence to measure the concurrence and purity of a two-qubit system. First tuned the coupler to decouple the qubits, then qubits Q1 and Q2 are prepared in one of the initial states: $|00\rangle$, $|01\rangle$,  $|10\rangle$, or $|11\rangle$. The qubits are next brought into resonance with TLS, $\delta$ represents the qubit frequency detuning from the TLS. After a variable delay, the two-qubit density matrix $\rho$ is measured using state tomography.
    \textbf{b,} Purity and \textbf{c,} Concurrence derived from state tomography. The top row of subplots shows the purity and concurrence extracted from measured density matrices for each initial state under different detuning $\delta$ and delays, while the bottom row presents the corresponding numerical simulation results. We observed that the purity and concurrence are asymmetric on both sides of $\delta = 0$, which is caused by frequency modulation errors on qubits. In the simulation, we set the detuning of Q2 relative to Q1 to 1 MHz, which best reproduces this effect. 
    \textbf{d,} Predicted fidelity of the two-qubit gate ($i$SWAP) as a function of the detuning $\delta$ between the qubit interaction resonance frequency and the TLS. The gate duration is fixed at $t_\mathrm{gate} = 60~\mathrm{ns}$, with TLS coherence times $T_1^\mathrm{TLS} = 44.7~\mu\mathrm{s}$ and $T_\phi^\mathrm{TLS} = 1.1~\mu\mathrm{s}$. Results are shown for different TLS-qubit coupling strengths and configurations where the TLS couples to one or both qubits.
    \textbf{e,} Predicted fidelity of the two-qubit gate versus the TLS relaxation rate $\Gamma_1^\mathrm{TLS} \equiv 1/T_1^\mathrm{TLS}$ for coupling strengths  $\tilde{g}_{1,2}/2\pi=1$~MHz and assuming zero TLS dephasing ($T_\phi^\mathrm{TLS} \to \infty$). The detuning $\delta$ is varied, and cases where the TLS couples to both qubits. 
    }
    \label{fig:error}
\end{figure*}


We experimentally investigate the TLS-mediated interaction between two idling qubits by examining their joint evolution under controlled conditions. The experimental protocol involves preparing both qubits in each of the four computational basis states, tuning them to the same frequency (detuned by $\delta$ from the TLS as shown in Fig.~\ref{fig:error}a), and maintaining their decoupling through careful adjustment of the tunable coupler (detailed in SM). Following variable evolution times, we reconstruct the two-qubit density matrix $\rho(t)$ through state tomography~\cite{niu2019learningnonmarkovianquantumnoise}, from which we calculate both the purity $P=\mathrm{tr}(\rho^2)$ and concurrence $\mathcal{C}$\cite{Wootters_concurrence_prl}(see SM), a simple metric that denotes degrees of entanglement. Results are presented in Fig.~\ref{fig:error}b-c.

In the absence of TLS coupling, the qubits would maintain their initial states indefinitely aside from relaxation. However, the presence of the TLS induces characteristic periodic excitation exchange between qubits and TLS that manifests as chevron patterns in the measured purity.  At resonance ($\delta=0$), in particular, a starting $\ket{11}$ state evolves through the sequence:
$\ket{11}\otimes\ket{0}^\mathrm{T} \to \frac{1}{\sqrt{2}}(|01\rangle + |10\rangle)\otimes\ket{1}^\mathrm{T} \to \ket{11}\otimes\ket{0}^\mathrm{T}$ 
(global phase ignored), where the superscript ``T'' denotes TLS. At time $t=1/(4\sqrt{2}\tilde{g})$, the two qubits reaches a maximally entangled state, a Bell state, while the TLS remains excited. Therefore, the presence of TLS can unintentionally induce additional entanglement between the two qubits.
This poses particular challenges for quantum error correction, as unwanted TLS excitations may accumulate and degrade subsequent operations, leading to temporally correlated errors.
Numerical simulations, shown in the bottom panels of Figs.~\ref{fig:error}b and c, successfully reproduce these results, validating our system modeling.

Building on our system characterization, we now investigate TLS impacts on quantum gate operations to quantitatively assess their contribution to gate errors. Numerical simulations of TLS-induced coherent errors for an $i$SWAP gate (see Supplemental Information Section XIV for simulation details and results for other gate types) are shown in Fig.~\ref{fig:error}d-e.

First, we find that gate errors are systematically higher when the TLS couples to both qubits compared to local coupling with just one qubit(Fig.~\ref{fig:error}d). The errors increase both when the qubit frequency approaches the TLS frequency ($\delta \to 0$) and when the qubit-TLS coupling strength increases. For a coupling strength of $10~\text{MHz}$, maintaining sub-$1\%$ error rates requires $\sim$300-500~MHz detuning from the TLS frequency, presenting significant challenges to frequency planning.

Second, we examine the influence of TLS dissipation ($\Gamma_1^\mathrm{TLS}$) on gate errors (Fig.~\ref{fig:error}e). When $\Gamma_1^\mathrm{TLS}$ is smaller than the coupling strength $\tilde{g}$, the gate error remains relatively constant and dominated by coherent effects. As $\Gamma_1^\mathrm{TLS}$ increases beyond $\tilde{g}$, the gate error exhibits non-monotonic behavior (except for the resonant case) --- first increasing before decreasing. 
These two regimes correspond to the anti-Zeno (underdamped) and Zeno (overdamped) effects\cite{greenfield2025unifiedpicturequantumzeno}, respectively. 

When $\Gamma_{\mathrm{TLS}} \gg \tilde{g}_{1,2}/2\pi$, the TLS relaxes too quickly to be excited, rendering it effectively "invisible" to the qubit. From a measurement perspective, strong TLS dissipation acts as a continuous measurement, corresponding to a quantum Zeno effect. As a result, the TLS cannot coherently exchange energy with the qubit, thereby suppressing TLS-induced two-qubit gate errors. The maximum of gate error emerges at $\Gamma_1^{\mathrm{TLS}} \approx \delta$.

\subsection{Discussions}

Our findings advance the understanding of defect interactions in state-of-the-art superconducting quantum processors. The identification of coupler-mediated TLS interactions reveals two previously overlooked phenomena: non-local coupling effects that extend beyond traditional qubit-local defect models, and dynamic coupling strengths that vary during processor operation. These effects impose stricter frequency allocation constraints, as shared defects demonstrate more severe gate performance degradation compared to local defects, and the tunable coupling strength ($\tilde{g} \propto 1/\Delta$) further complicates frequency planning.

The tunable coupler architecture presents unique opportunities for defect characterization. Intentional design of large coupler junctions can enhance measurable TLS density without compromising qubit coherence, creating a valuable testbed for comparing fabrication processes---particularly in high-quality materials where defect densities are naturally low.

The observed non-Markovian dephasing motivates more accurate modeling of TLS decoherence and its impact on qubit operations. Our noise spectroscopy protocol provides a standardized methodology to study defect-defect interactions, investigate microscopic origins of TLS, and evaluate mitigation strategies for next-generation quantum devices. The technique, together with other advancing identification approaches, establish superconducting quantum processors as powerful tools for both computation and materials science research.

\textit{Note added.} --- During the preparation of this manuscript, a recent work has shown that a TLS in the tunnel barrier of a transmon qubit can effectively couple to the readout resonator and thereby degrade the readout performance~\cite{lisenfeld2026readoutfailure}.

\begin{acknowledgments}
We thank Alex Opremcak, Mo Chen, J{\"u}rgen Lisenfeld, and Alexey Ustinov for helpful discussions. This work was supported by the Beijing Natural Science Foundation (Grants No. JQ25014), the National Natural Science Foundation of China (Grants No. 12322413, No. 92476206, and No. 92365206), and the Innovation Program for Quantum Science and Technology (Grant No. 2021ZD0301802)
\end{acknowledgments}

\bibliography{Biblio}
\end{document}